\documentclass[%
 reprint,
 amsmath,amssymb,
 aps,
]{revtex4-2}

\usepackage{graphicx}
\usepackage{dcolumn}
\usepackage{bm}

\usepackage[T1]{fontenc}
\usepackage{mathptmx}
\usepackage{mathrsfs}
\usepackage{physics}
\usepackage{simplewick}
\usepackage{mathrsfs}
\usepackage{amsthm,amsmath,amssymb}
\usepackage[bottom]{footmisc}
\usepackage{amssymb}
\usepackage{amsmath}
\usepackage{amsfonts}
\usepackage{bm}
\usepackage{graphicx}
\usepackage{subfigure}
\usepackage[dvipsnames]{xcolor}
\usepackage{calc}
\usepackage{epsfig}
\usepackage{color}
\usepackage[utf8]{inputenc}
\usepackage[colorlinks,citecolor=blue]{hyperref}
\usepackage{gensymb}

\newcommand{\be}{\begin{equation}}
\newcommand{\ee}{\end{equation}}
\newcommand{\bs}{\begin{split}}
\newcommand{\es}{\end{split}}

\begin{document}

\preprint{APS/123-QED}

\title{Quasi-monolithic Compact Interferometric Sensor Head Design with Laser Auto-alignment}

\author{Xiang Lin}
\author{Peng Qiu}
\author{Yurong Liang}
\author{Xiaofang Ren}
\author{Hao Yan}%
\email{yanhao@hust.edu.cn}
\author{Zebing Zhou}
\affiliation{Center for gravitational experiment, MOE Key Laboratory of Fundamental Physical Quantities Measurements, The School of Physics, Huazhong University of Science and Technology, Wuhan 430074, China}



\date{\today}

\begin{abstract}
Interferometers play a crucial role in high-precision displacement measurement such as gravitational-wave detection. 
Conventional interferometer designs require accurate laser alignment, including the laser pointing and the waist position, to maintain high interference contrast during motion. Although the corner reflector returns the reflected beam in parallel, there is still a problem of lateral beam shift which reduces the interference contrast. This paper presents a new compact interferometric sensor head design for measuring translations with auto-alignment. It works without laser beam alignment adjustment and maintains high interferometric contrast during arbitrary motion (tilts as well as lateral translation). Automatic alignment of the measuring beam with the reference beam is possible by means of a secondary reflection design with a corner reflector. A $20\times 10\times 10\,\mathrm{mm}^3$ all-glass quasi-monolithic sensor head is built based on UV adhesive bonding and tested by a piezoelectric (PZT) positioning stage. Our sensor head achieved a displacement sensitivity of $1\, \mathrm{pm}/\mathrm{Hz}^{1/2}$ at ${\rm 1\,Hz}$ with a tilt dynamic range over $\pm 200\,\mathrm{mrad}$. This optical design can be widely used for high-precision displacement measurement over a large tilt dynamic range, such as torsion balances and seismometers. 
\end{abstract}

\maketitle


\section{\label{sec:level1}INTRODUCTION}

Laser interferometers are widely used for high-precision displacement measurement in various fields such as gravitational wave detectors\,\cite{armano2021}, inertial sensors\,\cite{zhang2022quasi,hines2023compact,carter2020high,Yan_2022}, vibrometers\,\cite{Shang2020,yan2022all}, machine tools\,\cite{STEINMETZ199012}, Coordinate Measuring Machines (CMM)\,\cite{Umetsu_2005}, lithography\,\cite{Brueck2005}, etc. Besides, the ongoing space gravity detection missions Tianqin and Taiji require laser interferometer sensitivity of picometer and nanoradian level in the low-frequency bands of ${\rm 1\,mHz-1\,Hz}$\,\cite{Luo_2016,Li2020}. The need for such high-precision measurements places high demands on the design and construction of interferometers.

Various high-precision interferometers characterized by compact, stable, and wide dynamic range have been developed in recent decades. A comprehensive review of compact interferometry can be found in\,\cite{Watchi2018}. According to the different types of interference signals, interferometers can be divided into homodyne interferometers\,\cite{Lin2023}, heterodyne interferometers\,\cite{yan2020highly,yan2015dual,Shi2023,S2014Compact}, Fabry–Perot (FP) cavity interferometers\,\cite{Wu2023}, and modulation interferometers. Modulation interferometers can be subdivided into frequency modulation interferometers\,\cite{Gerberding:15,Isleif16,isleif2019compact}, phase modulation interferometers\,\cite{Heinzel10}, and amplitude modulation interferometers (including pulse or optical comb)\,\cite{Ye04}.

Among these interferometers, homodyne interferometers, which usually consist of a laser source, an optical bench, a target reflector and a DC photodetector, are preferred for their simplicity and compactness. To further improve the compactness and stability, optical sensor head designs based on Michelson-type or FP-type are usually chosen in practical applications and commercial interferometers, such as SmarAct, attocube IDS, quDIS, etc.\,\cite{Smetana2022,Smetana2023,Rohr23}. 

Compact interferometric sensor heads are important for the needs of large-range, high-precision displacement measurements in confined spaces, such as microstrain and micro-vibration monitoring and analysis\,\cite{Qu22,Wang2020}, and microsensor calibration testing. Moreover, compact miniature optical sensor heads have higher sensitivity potential with better opto-mechanical-thermal stability\,\cite{ressel2010ultrastable,preston2012quasi}. 

Conventional homodyne interferometers with plane target mirrors have some inherent drawbacks. Homodyne interferometers measure small displacement by reading changes in DC light intensity, which is also sensitive to the changes in interference contrast\,\cite{Watchi2018}. Precise laser alignment between the optical sensor head and the target mirror is required prior to use to achieve high interference contrast. In addition, high contrast needs to be maintained during motion, which limits the dynamic range of tilt motion. Finally, the tilt-to-length (TTL) coupling noise\,\cite{Wanner14,Hartig_2022}, i.e., the coupled response of tilt to translational light paths, depends on optical parameters such as beam waist, propagation distance, etc.

Using a corner reflector as a target instead of a plane reflector can solve the laser alignment problem during measurement to a certain extent, but it also has drawbacks\,\cite{yan2022all}. First, although the corner reflector enables the reflected beam to be returned parallel, there is still a problem of lateral beam shift which also reduces the interference contrast\,\cite{Xia:23}. Second, if the measurement beam covers the center of the corner reflector, the returned laser wavefront will split into six flaps (depending on the three dihedral-angle errors of the corner reflector), causing displacement coupling errors.

Many efforts have been made to improve the angular dynamic range of interferometers. A cat’s eye retroreflector can be used to increase the tilt tolerance range in interferometry\,\cite{Pena-Arellano}. A target corner reflector combined with a fixed reference plane mirror allows interferometers to measure a translation over a large angular dynamic range\,\cite{Zhang2007}.

In this paper, we proposed a novel Michelson-type interferometric sensor head design for measuring translations with laser auto-alignment. Combined with a corner reflector, this design automatically guarantees a high interference contrast, greatly increasing the dynamic range while avoiding coupling errors in other degrees of freedom of motion, including tilts and lateral translations. A $20\times 10\times 10\,\mathrm{mm}^3$ all-glass quasi-monolithic compact optical sensor head is built and tested. To improve compactness and stability, we use UV-adhesive bonding technology to glue the optical components\,\cite{Lin2023}. The principle and the optical design of the sensor head are detailed and the experimental setup and sensitivity results are given in the following sections.

This paper is organized as follows. In Section II, we briefly outline the measurement principle and experimental layout of the experimental setup. Section III is devoted to the experimental results of dynamic range and displacement measurement sensitivity of the homodyne interferometer. The summary and further discussions are presented in the last section.

\section{\label{sec:level2}DESIGN and WORKING PRINCIPLE}

\subsection{\label{sec:level2.1}Optical Design}

\begin{figure}[h]
\includegraphics[width=0.5\textwidth]{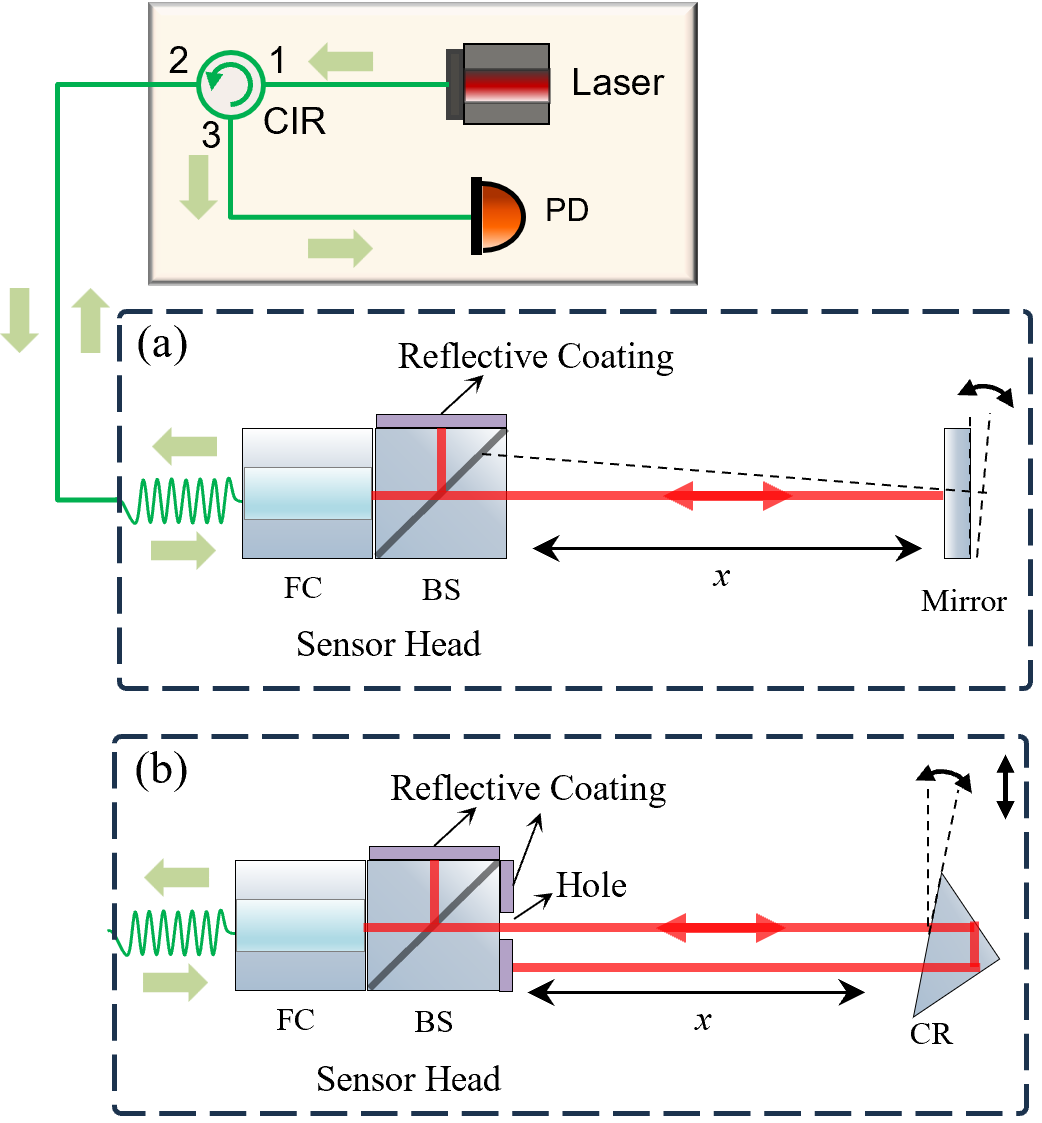}
\caption{\label{fig1} Optical schematic of the Michelson-type interferometer with a compact optical sensor head for the translation measurement. A laser beam from the laser source is injected into a circulator (CIR) via a polarization-maintaining fiber at port 1, and the output port 2 is connected to the optical sensor head via a fiber collimator (FC). The homodyne interference signal coming back from the optical sensor head enters the photodetector (PD) through port 3 of the circulator. Panel (a) shows a typical Michelson-type sensor head with a target plane reflector. The sensor head consists of a fiber collimator and a custom beam splitter (BS) with reflective coatings on the upper side. The collimated beam is split into two beams at a 50/50 ratio by the BS and reflected back to the FC by the target plane mirror. Panel (b) shows the new optical sensor head design with a target corner reflector. The custom beam splitter (BS) has reflective coatings on the two perpendicular side surfaces. A small coating hole is left in the direction of the sensitive axis for the laser to enter and exit. The measurement laser is reflected both by the corner reflector (CR) and the reflective coating.}
\end{figure}

The optical schematic of the homodyne interferometer with a compact optical sensor head for the translation measurement is shown in Fig.\,\ref{fig1}. The interferometric system mainly consists of five components: a laser source, a photodetector (PD), a circulator (CIR), a customized optical sensor head, and a target corner reflector (CR). The optical sensor head comprises a fiber collimator(FC) and a custom beam splitter (BS) with reflective coatings on side surfaces. The two optical elements (FC and BS) are glued together to form a quasi-monolithic compact sensor head, as shown in the figure. For a typical Michelson-type sensor head, the measurement laser beam is reflected directly back to the sensor head by the target plane reflector, as shown in Fig.\,\ref{fig1}(a). This design, in order to maintain high interference contrast, requires a high level of alignment in the motion of the target plane mirror\,\cite{Gerberding2021,Smetana2023}. 

Laser self-collimation can be achieved if an additional coating is added to the front side of the BS and a target corner reflector is used instead of the plane mirror, as shown in Fig.\,\ref{fig1}(b). The new design works as follows: First, the laser is injected into the optical sensor head via a circulator from port 1 to port 2. Second, the laser beam from the fiber collimator is split by the beam splitter inside the optical sensor head. The reflected reference laser beam is reflected back to the fiber collimator by the upper side surface with a reflective coating and the beam splitter, and the measurement laser beam in the direction of the sensitive axis passes through the coating hole. Third, the measurement laser beam from the sensor head hits the targe corner reflector and is reflected back to the sensor head in parallel with a small lateral shift, as shown in the figure. Fourth, the lateral shift beam is reflected back to the corner reflector and the optical sensor head in the same path by the reflective coating surface of the beam splitter from outside. The two reflected beams (the reference beam and measurement beam) interfere at the beam splitter and return to the inside of the fiber. Finally, the optical interference signal is detected by the photodetector via the circulator from port 2 to port 3. 

It is worth noting that our new optical sensor head can also work in the same way as the typical Michelson-type sensor head design shown in Fig.\,\ref{fig1}(a), reflecting once with a target plane mirror.

The new optical sensor head has special geometric requirements on the surfaces containing reflective coatings. The reflected laser beams (reference and measurement laser beams) must return to the inside of the fiber to maintain a high interference contrast. The reference laser beam reflected by the beam splitter must be perpendicular to the upper reflective coating surface to return along the original path. The transmitted laser beam reflected by the corner reflector with a lateral shift must also be perpendicular to the sensitive-axis reflective coating surface from outside to return along the original path. To achieve this, first, the two coated surfaces of the beam splitter need to be perpendicular to each other; then, the relative attitudes of the fiber collimator and the beam splitter need to be carefully adjusted when bonding the optical sensor head.

Once the new optical sensor head has been fabricated, the measurement laser beam automatically returns to the sensor head and interferes with the reference laser beam with high contrast. In Fig.\,\ref{fig1}(b), the parallel return beam is deflected downward due to the center of the corner reflector is below the optical axis; if the corner reflector moves upward, the parallel return beam will be deflected upward laterally. In both cases, it is guaranteed that the measurement laser beam will be reflected twice and return to the fiber. The dynamic range of motion of the target corner reflector is limited by the lateral displacement, as shown in Fig.\,\ref{fig1}(b). The returning measurement laser beam can not extend beyond the surface of the reflective coating. This requires that the center of the target corner reflector not deviate too far from the exit optical axis.

Notably, this sensor head design responds only to translational motion along the sensitive axis, thus effectively suppressing coupling errors arising from motion in other degrees of freedom (e.g. TTL coupling)\,\cite{Yan_2022,Wanner14,Hartig_2022}.

\subsection{\label{sec:level2.2}Measurement Principle}
The measurement laser beam from the target corner reflector and the reference laser beam inside the optical sensor head interference in the beam splitter and detected by the photodetector. Assuming the homodyne interference signal be written as:
\begin{equation}
    P=P_0[1+C\cdot\cos{\left(\varphi_m-\varphi_r\right)}],
\label{eq1}
\end{equation}
where $P$ is the laser power on the photodiode, $P_0$ is the average laser power on the photodiode, $C$ is the interference contrast. $\varphi_m$ and $\varphi_r$ are the cumulative path phases of the measurement beam and the reference beam, respectively. The interference contrast is given by the equation
\begin{equation}
    C=\frac{P_{max}-P_{min}}{P_{max}+P_{min}},
\label{eq2}
\end{equation}
where $P_{max}$ and $P_{min}$ are the maximum power and minimum power, respectively. 

The interference contrast $C$ depends mainly on the beam-waist ratio, the propagation distance, the wavefront distribution, the center position offset, and the pointing deviation of the two interfering laser beams. In this optical sensor head design, the interference contrast can be maintained at a high level. The actual values of the average laser power $P_0$ and the interference contrast $C$ can be experimentally calibrated by moving the target reflector over a wide range of translations. The interference signal light intensity change is monitored and then the phase change is calculated by Eq.\,(\ref{eq1}).

Considering that the measurement laser beam is reflected twice between the optical sensor head and the target corner reflector (as shown in Fig.\,\ref{fig1}, the target translation $x$ is given by the equation
\begin{equation}
    x=\frac{\lambda}{8\pi}(\varphi_m-\varphi_r),
\label{eq3}
\end{equation}
where $\lambda$ is the laser wavelength. The above Eq.\,(\ref{eq1}-\ref{eq3}) is a homodyne interference readout, and the measurement signal is a DC signal. We can also use laser frequency modulation technique to convert it into AC signals\,\cite{Gerberding:15}. 

Note the special case in Fig.\,\ref{fig1}(b), where the measurement beam reflected for the first time from the corner reflector returns directly to the sensor head along the small hole. In this case, there is only one reflection here, similar to Fig.\,\ref{fig1}(a). The target translation $x$ is given by the equation
\begin{equation}
    x=\frac{\lambda}{4\pi}(\varphi_m-\varphi_r). 
\label{eq4}
\end{equation}
Considering the lateral displacement coupling of the reflected beam from the corner reflector, we should try to avoid this situation.


\section{\label{sec:level3} Prototype and 
 TEST RESULTS}
\subsection{\label{sec:level3.1}Prototype setup}
\begin{figure}[h]
\includegraphics[width=0.48\textwidth]{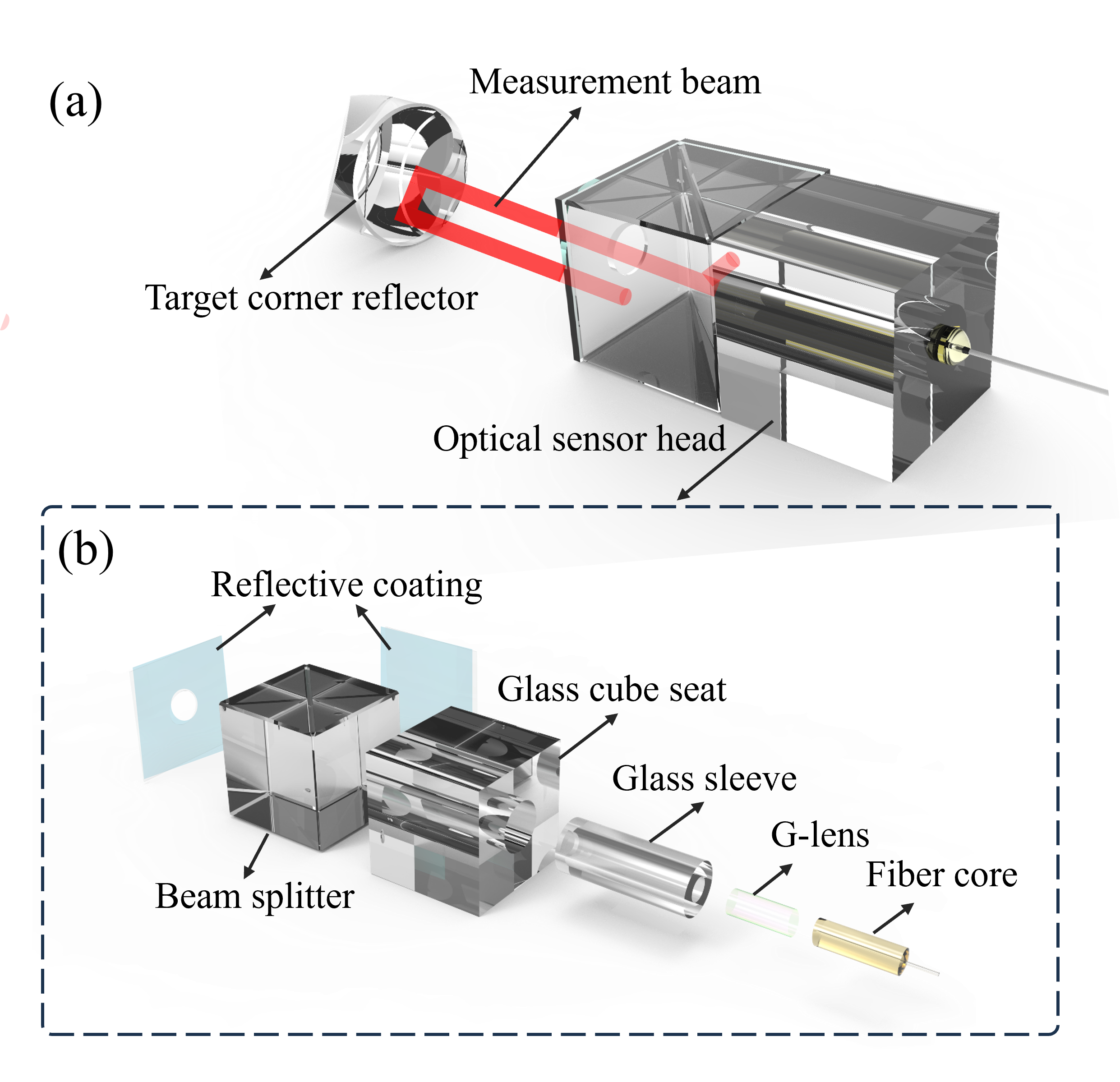}
\caption{\label{fig2}Three-dimensional quasi-monolithic design of the customized optical sensor head and the target corner reflector. Panel (a) shows the optical sensor head and the target corner reflector. The measurement laser beam (red) is first reflected with a lateral shift by the corner reflector and then reflected back in the same path by the reflective coating of the sensor head. Panel (b) shows the detail of the optical sensor head, including the fiber core, G-lens, glass sleeve, glass cube seat, and cube beam splitter with two reflective coating surfaces. All components are glued together with dimensions of $20\times 10\times 10\,\mathrm{mm}^3$.}
\end{figure}

In order to obtain high stability and precision, we built an all-glass quasi-monolithic compact optical sensor head via UV-adhesive bonding\,\cite{Lin2023}. The three-dimensional quasi-monolithic optical sensor head design is shown in Fig.\,\ref{fig2}. The measurement laser beam exits through the small reflective coating hole in the front surface of the beam splitter. The laterally shifted returned beam from the target corner reflector is reflected back in the same path by the reflective coating of the sensor head. Finally, the interference signal returns along the laser injection fiber and is detected by a photodetector. The all-glass quasi-monolithic optical sensor head consists of a fiber core, a G-lens, a glass sleeve, a glass cube seat, and a cube beam splitter with two reflective coating surfaces, as shown the Fig.\,\ref{fig2}(b). All these components are bonded together via UV adhesive to form a rectangular compact sensor head with dimensions of $20\times 10\times 10\,\mathrm{mm}^3$.

\begin{figure}[h]
\includegraphics[width=0.48\textwidth]{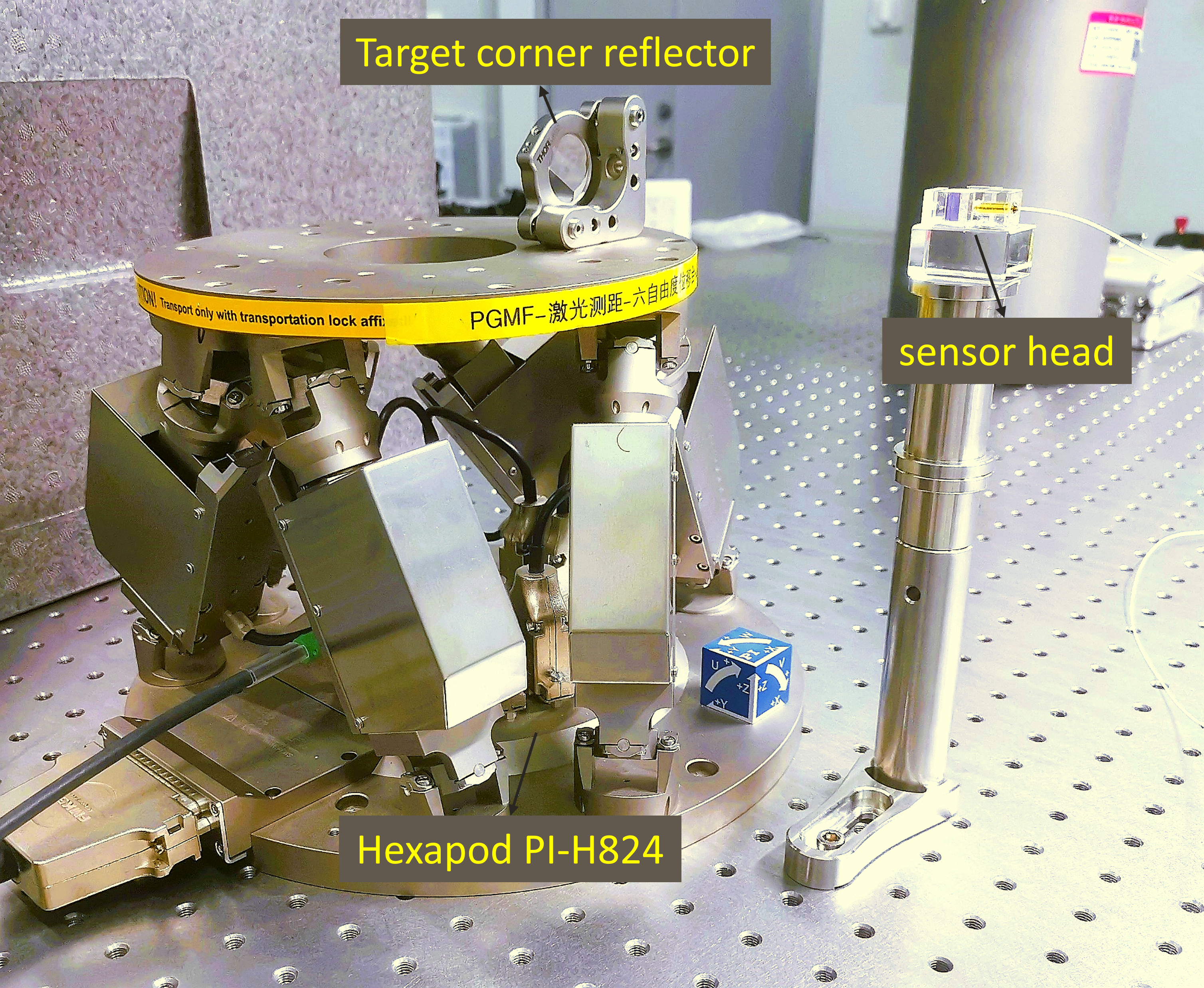}
\caption{\label{fig3} Photo of the experimental prototype setup. The target corner reflector (UnionOptic CNP0025) is mounted to the top of the Hexapod (PI-824) through a high-stability mirror mount (Thorlabs POLARIS-K1E2), and the quasi-monolithic optical sensor head is bonded to the top of a metal cylinder. The Hexapod is controlled by a computer to scan step by step around the z-axis.}
\end{figure}

The experimental prototype setup is shown in Fig.\,\ref{fig3}. The quasi-monolithic optical sensor head is bonded to the top of a metal cylinder. To test the interferometer system's tilt dynamic range, the target corner reflector is mounted on the upper plate of the Hexapod through a mirror mount. A laser (Nd:YAG) with a wavelength of 1064 nm and a power of 1 mW is connected to a fiber optic circulator (Throlab CIR1064PM). The laser emitted by the circulator is injected into the quasi-monolithic sensor head via a fiber. The key parameters of the experimental setup are given in Tab.\,\ref{tab1}.

\begin{table}
\caption{\label{tab1}The key parameters of the experimental setup.}
\begin{ruledtabular}
\begin{tabular}{cccccccc}
 Parameters&Value\\
\hline
laser wavelength& $1064\,\mathrm{nm}$\\
beam waist diameter& $\sim 1\,\mathrm{mm}$\\
input beam power& $\sim 1\,\mathrm{mW}$\\
size of the corner reflector (diameter)& $25.4\,\mathrm{mm}$\\
size of the optical sensor head & $20\times 10\times 10\,\mathrm{mm}^3$\\
material of the optical sensor head & fused silica\\
optical bonding method & UV adhesive gluing\\
\end{tabular}
\end{ruledtabular}
\end{table}

Adjust the position of the target corner cube reflector to ensure that the reflected measurement beam with a lateral displacement does not exceed the reflective coating area of the beam splitter. The center of the corner cone must not be too far away from the measuring optical axis. The motion of the hexapod is controlled by a computer.

\subsection{\label{sec:level3.2}Experiment Results}
\subsubsection{\label{sec:level3.2.1}Contrast and Dynamic Range}

Dynamic range is one of the key parameters of high-precision displacement sensors. The sensitive translation dynamic range of interferometers is usually not a problem. It is mainly limited by the collimation range of the laser or the way the interference phase is demodulated. However, the tilt dynamic range of interferometers with plane reflector is usually limited by the interference contrast to about $mrad$ level. Beyond this range, the contrast of the interferometer decreases rapidly. 

There are two methods to measure the contrast of a homodyne interferometer. The first direct method is to drive the target reflector to scan over a wide range of translation motion (more than half a wavelength), covering at least one bright fringe and one dark fringe. The second method is deep frequency modulation of the laser. A phase modulation signal will be generated in the homodyne interferometer with unequal arm lengths\,\cite{Gerberding:15,isleif2019compact}. The phase modulation depth is proportional to the unequal arm lengths and the frequency modulation amplitude. We use the second method in our interference contrast tests.

The interference contrast of the quasi-monolithic sensing probe needs to be tested first. We modulate the laser frequency at $\omega_m=2\pi\cdot 10\, \mathrm{Hz}$ and amplitude of $A_m\approx0.25\,\mathrm{GHz}$ with an internal piezoelectric transducer (PZT) actuator. The power observed at the photodetector is given by the equation\,\cite{Smetana2022}
\begin{equation}
P(t)=P_0\left[1+C\cos{\left(\varphi_m-\varphi_r\right)}+m\cos{\left(\omega_mt\right)}\right],
\label{eq5}
\end{equation}
where $m=8\pi A_m \Delta L/c$ is the unitless modulation index, $\Delta L$ is the longitudinal imbalance of the Michelson arms and $c$ is the speed of light. The measurement signal passes through a high-gain silicon-based photodetector, and the raw photoelectric signal is acquired by a digital acquisition device (Liquid Instuments Moku:Lab).

\begin{figure}[h]
\includegraphics[width=0.48\textwidth]{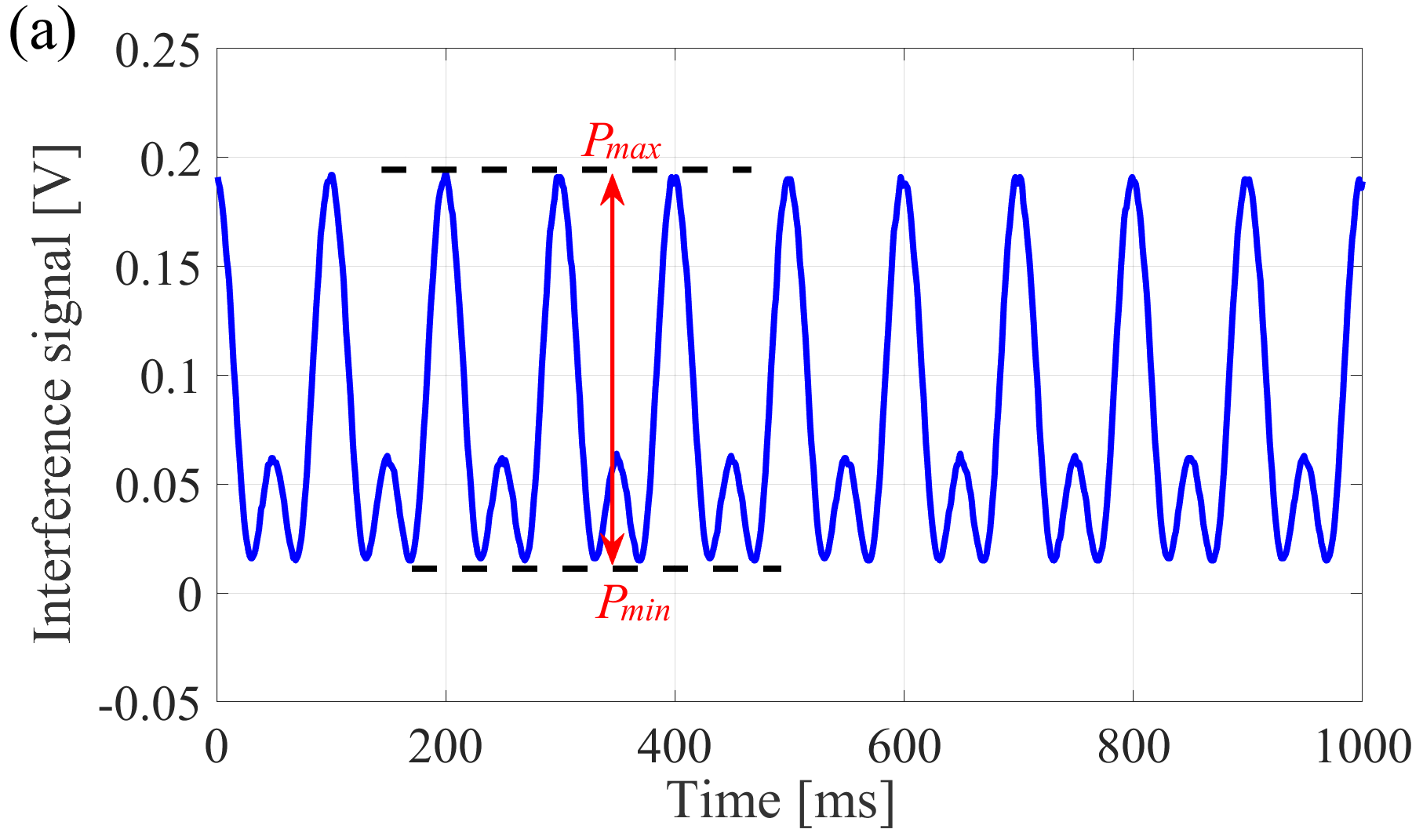}
\includegraphics[width=0.48\textwidth]{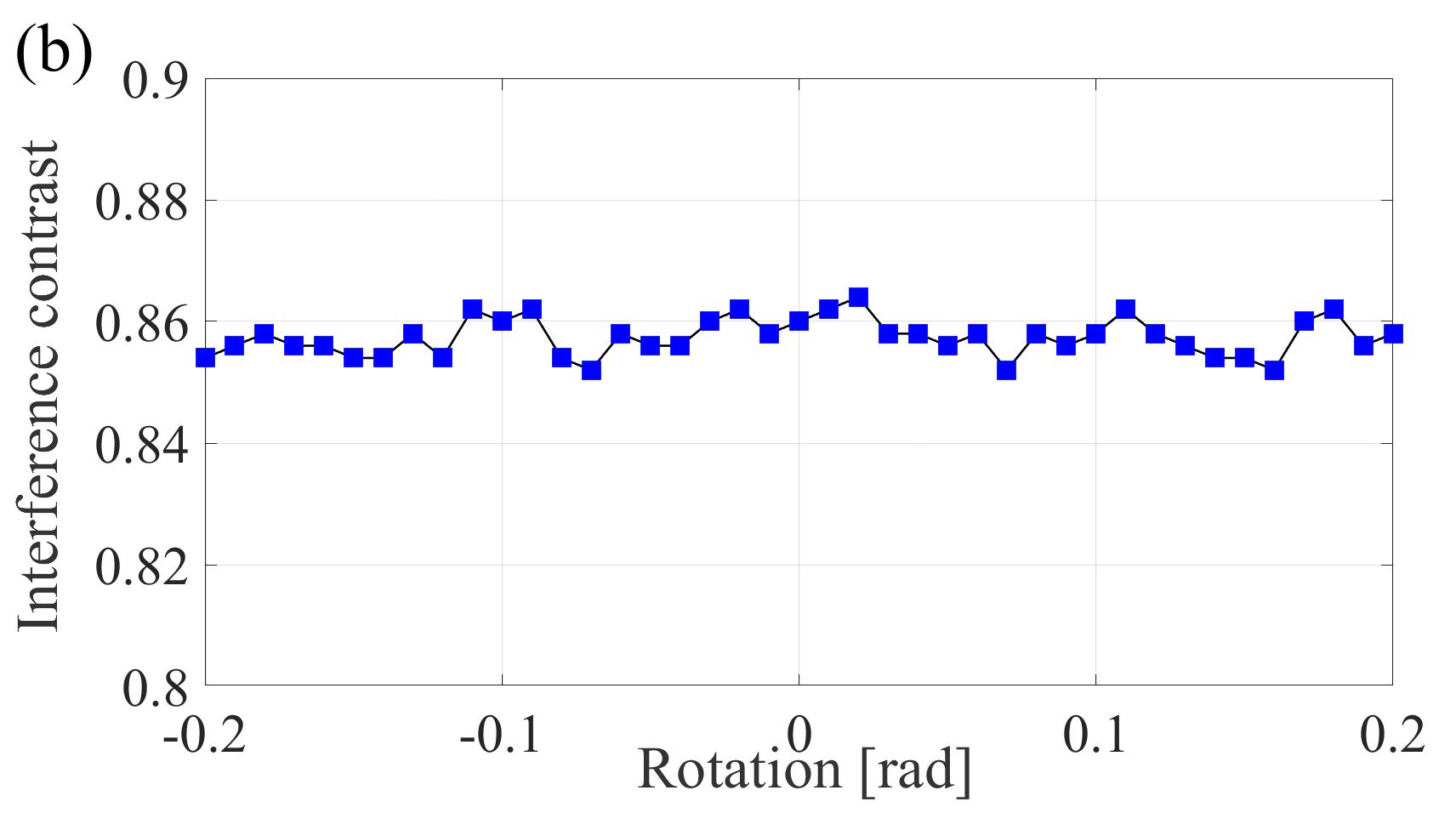}
\includegraphics[width=0.48\textwidth]{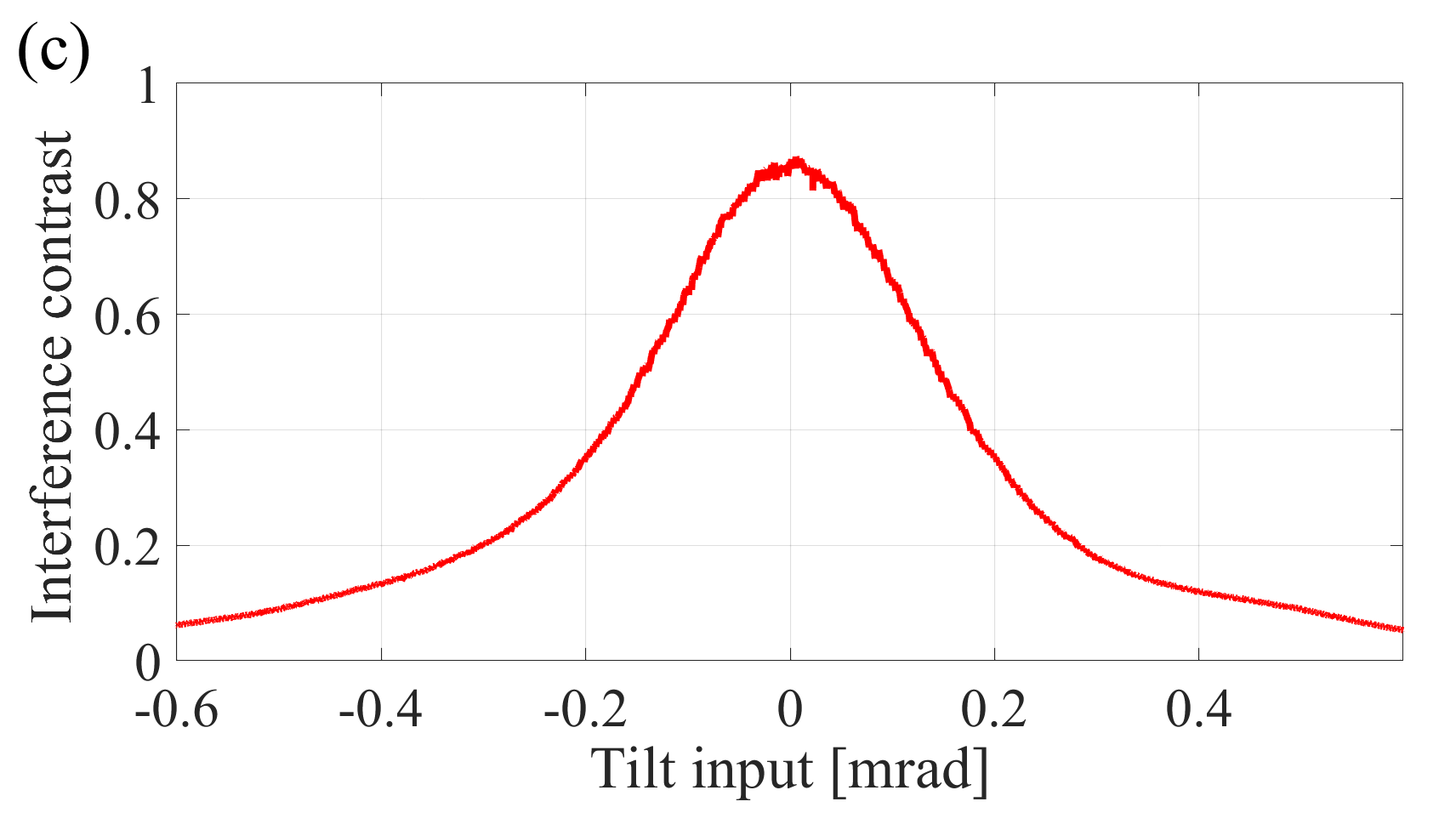}
\caption{\label{fig4} Interference contrast and tilt dynamic range test results. (a) The interference signal of the setup with the laser frequency modulated. (b) The curve of the interference contrast of our new auto-aligning optical sensor head with the target corner reflector rotation. The angular dynamic range is scan from $-200\,\mathrm{mrad}$ to $+200\,\mathrm{mrad}$ in steps of $10\,\mathrm{mrad}$. (c) The curve of the interference contrast of the same optical sensor head with a plane mirror rotation (a typical Michelson homodyne interferometer design, just the same as the sensor head of the SmarAct company\,\cite{Smetana2022}). 
The angular dynamic range of this Michelson-type sensor head is about $\pm0.2\,\mathrm{mrad}$ ($C<0.4$).}
\end{figure}

The unequal arm length $\Delta L$ is set to about $0.1\,\mathrm{m}$ and the unitless modulation index {$m\approx2 > \pi/2$}. Then the interference contrast can be resolved by Eq.\,\ref{eq2} from the modulated interference signal shown in Fig.\,\ref{fig4}(a). The initial homodyne interferometer is well aligned and the contrast is about $C\approx0.86$ (Fig.\,\ref{fig4}(a)). To test the tilt dynamic range of our quasi-monolithic sensing head, we utilized a Hexapod (PI-H824) to rotate the yaw angle of the target angular reflector, as shown in Fig.\,\ref{fig3}. The Hexapod we used has a limit rotation range of $\pm 12.5\degree$ (around the z-axis), or$\pm 220\,\mathrm{mrad}$. Therefore, the angle scan range of the contrast test is set from $-200\,\mathrm{mrad}$ to $+200\,\mathrm{mrad}$ in steps of $10\,\mathrm{mrad}$. 

The test results of the contrast variation with the corner reflector rotation are shown in Fig.\,\ref{fig4}(b). The interference signal maintains high contrast ($C>0.85$) over a rotation range of $\pm 200 \mathrm{mrad}$. The slight fluctuations in the contrast curve are mainly due to the unevenness of the reflective coating. 

For comparison, the same test was done using a plane mirror with the same sensor head. In this case, the measurement laser beam reflected from the plane mirror returns directly to the fiber in the same path, as shown in Fig.\,\ref{fig1}(a). The results are shown in Fig.\,\ref{fig4}(c). The interference contrast decreases rapidly with the rotation of the target plane mirror, falling below $0.4$ beyond a rotation range of $\pm 0.2 \mathrm{mrad}$. 

It is worth noting that our auto-aligning optical sensor head has the potential to achieve a greater angular dynamic range based on the design in Fig.\,\ref{fig1}. Even so, the results of this angular dynamic range still far exceed the $\sim\,\mathrm{mrad}$ level of typical interferometers\,\cite{Wanner14}.

\subsubsection{\label{sec:level3.2.2}Resolution and sensitivity}
To determine the resolution of the new optical sensor head, a piezoelectric transducer (PZT) actuator is employed. The voltage signal from the signal generator (Keysight, 33622A) drives the mirror movement (translation or tilts) through the PZT.

\begin{figure}[h]
\includegraphics[width=0.5\textwidth]{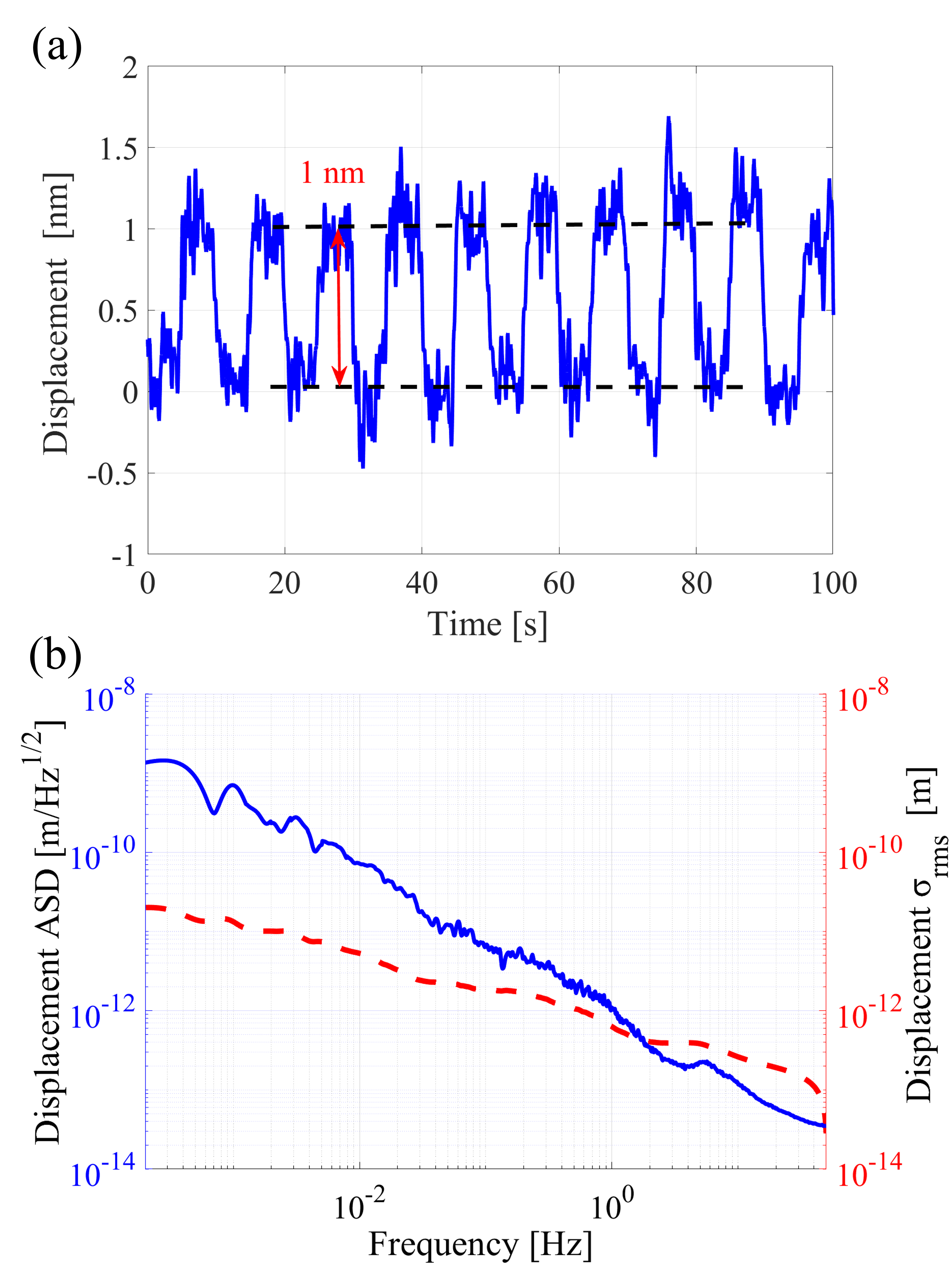}
\caption{\label{fig5} Test results of the resolution and sensitivity. Panel (a) shows a translation displacement signal driven by the PZT (Thorlabs), a
square wave with an amplitude of $1\,\mathrm{nm}$ and a period of $5\,\mathrm{s}$. Panel (b) shows the displacement sensitivity curve of the sensor head. The blue solid curve is the logarithmic amplitude spectral density (ASD), and the red dashed curve is the root mean square error.}
\end{figure}

The test results are shown in Fig.\,\ref{fig5}. In the resolution test, the target corner reflector is driven by a PZT with a small translation. The voltage signal driving the PZT is a square wave with an amplitude of 
$10\,\mathrm{mV}$ and a period of $5\,\mathrm{s}$, corresponding to a displacement of about $1\,\mathrm{nm}$ based on a coefficient of $100\,\mathrm{nm/V}$. The signal output in time domain is shown in Fig.\,\ref{fig5}(a). To further measure the sensitivity, the driven voltage is switched off. The sensitivity curve is shown in Fig.\,\ref{fig5}(b). A sensitivity of $1 \mathrm{pm/Hz^{1/2}}$ at $1\,\mathrm{Hz}$ is achieved as shown in the blue solid curve. 

The root-mean-square (rms) error between frequency $f$ and $fs/2$ can be calculated from the PSD (power spectral density) by
\begin{equation}
\sigma_{rms}(f)=\sqrt{\int_{f}^{fs/2}{\rm ASD}^2(f')\cdot df’},
\label{eq6}
\end{equation}
where $fs$ is the sampling frequency, ${\rm ASD}(f')$ is the amplitude spectral density. According to the Nyquist–Shannon sampling theorem, the upper limit of the PSD integral is $fs/2$. The red dashed curve in Fig.\,\ref{fig5}(b) shows
the root-mean-square (rms) error of the interferometer system. A displacement resolution of picometer level is achieved.

\section{\label{sec:level4}DISCUSSION AND CONCLUSIONS}
In this paper, a novel Michelson-type interferometric sensor head design for translation measurement with laser auto-alignment is proposed. This optical sensor head design, combined with a target corner reflector, automatically guarantees a high interference contrast. Compared to conventional interferometers using a target plane reflector, this optical design greatly increases the dynamic range while avoiding coupling errors in other degrees of freedom of motion, including rotations and lateral translations. To verify the performance of our design, an all-glass quasi-monolithic compact sensor head was built with UV-adhesive bonding technology and tested by a Hexapod and PZT stages.

The performance of the experimental prototype setup is tested. The experimental results show that a high contrast ratio above $85\%$ is maintained over the dynamic range of $\pm200\,\mathrm{mrad}$, and a sensitivity of $1 \mathrm{pm/Hz^{1/2}}$ at $1\,\mathrm{Hz}$ is achieved.
The dynamic range of the new design is more than 1,000 times higher than that of a conventional Michelson-type design with the same optical parameters. It is worth noting that our auto-aligning optical sensor head
has the potential to achieve a greater angular dynamic range.


This auto-alignment interferometric sensor head design with a large dynamic range and high precision can be widely used for future multiple degrees-of-freedom displacement metrology and applications, such as torsion balances and seismometers.


\begin{acknowledgments}
We thank members of Prof. Yiqiu Ma for useful discussions and Dr. Boping Chen of Tel Aviv University for his valuable feedback. The authors acknowledge the experimental facility support from the PGMF (National Precision Gravity Measurement Facility), National Key Research and Development Program of China (2022YFC2203901), National Natural Science Foundation of China (12105375).
\end{acknowledgments}

\bibliography{ref}

\nocite{*}

\end{document}